# Information acquisition in Adapt/Exchange decisions: When do people check alternative solution principles?

Romy Müller & Maria Pohl

*Faculty of Psychology, Chair of Engineering Psychology and Applied Cognitive Research, TUD Dresden University of Technology, Dresden, Germany*

Corresponding author:

Romy Müller

Chair of Engineering Psychology and Applied Cognitive Research

TUD Dresden University of Technology

Helmholtzstraße 10, 01069 Dresden, Germany

Email: romy.mueller@tu-dresden.de

Phone: +49 351 46335330

Fax: +49 351 46337741

ORCID: 0000-0003-4750-7952

**Abstract**

Many problems can be solved in two ways: either by adapting an existing solution, or by exchanging it for a new one. To investigate under what conditions people consider new solutions, we traced their information acquisition processes in a simulated mechanical engineering task. Within a multi-step optimisation procedure, participants could either adapt the properties of a currently used machine component, or exchange this component for a new one. They had the opportunity to check whether the solutions met a set of requirements, which was varied systematically. We investigated whether participants would consistently check both solutions, or whether they would satisfice, ignoring the new solution as long as the current one was good enough. The results clearly refuted consistent checking, but only partly confirmed satisficing. On the one hand, participants indeed checked the new solution least often when the current one was applicable without problems. On the other hand, in this case the new solution still was not fully ignored. However, the latter finding could be traced back to a few participants who diverged from our anticipated strategy of first checking the current solution, and directly went for the new one. The results suggest that in Adapt/Exchange decisions, people do not usually check both solutions in an unbiased manner, but rely on existing solutions as long as they are good enough.

# 1. Introduction

How do problem solvers decide between adapting a current solution and exchanging it for a new one? While the trade-off between stability and flexibility has received ample empirical attention in cognitive psychology (Chrysikou et al., 2014; Geddert & Egner, 2022; Goschke, 2013; Hommel, 2015), it is unclear what these theories and findings can tell us about complex problem solving. In complex systems, people often face ill-defined problems: the situation, the goal, and the ways to achieve this goal are underspecified (Dörner & Funke, 2017; Hollnagel, 2012; Perrow, 1984). Accordingly, decision makers need to collect information about the system, predict the outcomes of interventions, check different solution alternatives, and use their creativity to put these solutions into practice. However, an exhaustive exploration of all potential solutions is rarely possible. Thus, problem solvers often employ heuristics (Fischer et al., 2012) and show a tendency economise by relying on less resource-demanding strategies (Schoppek, 2023). Moreover, real-world problem solving typically is not a one-time activity, and thus previous solutions to similar problems are often available. If these solutions still produce satisfactory outcomes, they might simply be re-applied (perhaps with some modifications), instead of generating a new solution from scratch. Given problem solvers' tendency to economise, it is questionable whether they will flexibly consider alternative solutions when it is not mandatory.

## 1.1 Stability versus flexibility in decision making and Adapt/Exchange decisions

Psychological research on decision making provides ample evidence that people indeed favour stability over flexibility, and often stick with previous choices. Such status quo bias (Samuelson & Zeckhauser, 1988) or choice perseveration (Senftleben et al., 2019) has been observed in various decision contexts (Betsch et al., 2002; Dutt & Gonzalez, 2012; Erev et al., 2010; Rakow & Miler, 2009; Scherbaum et al., 2013). Similarly, people exhibit a clear preference to go with defaults, particularly when these defaults reflect the status quo (Jachimowicz et al., 2019). This can sometimes have far-reaching consequences, for instance when deciding about organ donations (Johnson & Goldstein, 2003) or pro-environmental behaviour (Pichert & Katsikopoulos, 2008). However, the stability-flexibility trade-off is quite simple in most psychological studies: participants merely have to decide whether to make a change or not. Conversely, many real-world problems require decision makers to choose between different types of changes: local versus global ones. That is, when an existing solution no longer produces satisfactory outcomes, you can either modify its details (Adapt), or apply a fundamentally different solution principle (Exchange) (Müller & Urbas, 2017). Such Adapt/Exchange decisions also are a cornerstone of problem solving in many vocational settings. For instance, when mechanical engineers re-design a machine, they can either adapt the properties of a component that is currently used (e.g., change its dimensions and material) or replace it by a completely new component that fulfils the same function but might overcome the shortcomings of the current component.

Adapt/Exchange decisions differ from the decisions typically studied by psychologists in theoretically interesting ways (Müller & Urbas, 2020). In short, they reflect decisions between solving a problem by means of local or global changes. These two options are qualitatively different in the sense that only some of their attributes are shared. Given that Adapt and Exchange are different means of achieving an abstract goal, decision making is not merely a matter of preference. Yet, choosing one of these solutions does not fully determine the outcome, as this outcome critically depends on how the change is implemented. Due to these peculiarities of Adapt/Exchange decisions, their underlying cognitive and behavioural processes are poorly understood. Particularly, do decision makers consider both solution principles equally, and thoroughly gather all relevant information about them before making their choice? A previous study suggests otherwise (Müller & Urbas, 2020): in a chemical process control scenario, participants seemed to refrain from checking the consequences of exchanging the solution as long as adapting the current solution was good enough. Such satisficing (Simon, 1956) was inferred from two findings. First, the frequency of Exchange choices depended on the costs of Adapt, instead of on the cost ratio between Adapt and Exchange. Second, Adapt choices were much faster than Exchange choices, suggesting that the latter included an additional cognitive process that was otherwise omitted (i.e., checking the alternative solution). Admittedly, these performance outcomes only provide very indirect evidence of satisficing, and cannot actually tell us whether participants indeed ignored the alternative solution. To obtain more conclusive evidence, the present study used an experimental setup that allows us to trace participants' information acquisition processes.

Under what conditions do people check alternative solutions when making Adapt/Exchange decisions? This was investigated in a simplified mechanical engineering task, in which participants chose between adapting the properties of a currently used machine component and exchanging it for a new component. They were free to examine the quality of one or both solutions by performing a sequence of cognitively demanding checks. We hypothesised that participants would refrain from checking the alternative solution when a check of the current solution revealed that the latter was good enough. This satisficing was contrasted with the alternative hypothesis that participants will consistently check both solutions. Before describing the experimental setting, we will elaborate on the nature of Adapt/Exchange decisions, and discuss to what extent their information acquisition processes can be inferred from previous research in various decision contexts.

## 1.2 Decisions within an abstraction hierarchy

The essence of Adapt/Exchange decisions is a choice between solving problems on a low level of abstraction versus moving up in the abstraction hierarchy. To understand this difference, one must consider that complex systems can be conceptualised in *abstraction hierarchies* (Naikar, 2017; Rasmussen, 1985): one and the same system can be described with regard to either its goals, the functions carried out to achieve these goals, the components realising these functions, or the properties characterising these components. Higher levels of abstraction define *why* the contents of a given level are needed, whereas lower levels define *how* they are implemented. Figure 1 illustrates this principle by using the example of a packaging machine. The goal of such a machine might be to maximise the output of products packaged per time unit. This goal can be achieved by combining different abstract functions or product handling tasks, such as transporting, filling, and forming (Bleisch et al., 2011). Each of them can be specified further, with the number of functional levels depending on the goal of description and the type of system (Vicente & Rasmussen, 1992). For instance, transporting can mean that the product is turned over. Ultimately, functions are realised by physical machine components. In the case of turning over, the component could be a turning wheel or a turning belt. These components have properties, such as their dimensions and the materials they are made of.

Successful problem solving requires a *flexible navigation between these abstraction levels*, instead of getting stuck on a particular one (Hall et al., 2006; Janzen & Vicente, 1998). Especially when problems are complex, it is beneficial to move up in the abstraction hierarchy (Ham & Yoon, 2001). This is because higher levels of abstraction allow problem solvers to find suitable alternatives when it is impossible to solve a problem by merely modifying the details of the current solution (Hajdukiewicz & Vicente, 2002). However, people may not always be so flexible: instead of considering an exchange of the solution principle, they often try to adapt it by merely making minor modifications. This lack of fundamental changes can be observed in various problem solving contexts.

One example is the well-known phenomenon of *design fixation* (Alipour et al., 2018; Youmans & Arciszewski, 2014). It occurs when people remain focused on a product's physical implementation, while being unable to revisit the more general ideas underlying this implementation (Jansson & Smith, 1991). In other words, designers remain stuck in a so-called configuration space, instead of moving to the more abstract concept space, although real change can rarely occur without moving to concept space. Previous research has investigated forms of design fixation that are more conceptual or more knowledge-based, and various cognitive mechanisms have been elaborated, such as unconscious adherence, conscious blocking, or intentional resistance (for an overview see Youmans & Arciszewski, 2014). However, a common theme across design fixation studies is that people face difficulties in recognising or generating new solutions.

In contrast, we investigate *processes of decision making* in Adapt/Exchange scenarios: we focus on situations in which people are fully aware of the possibility of using an alternative solution. Instead of asking whether they come up with this solution all by themselves, we ask whether they choose to seek out information about this solution, given that both the solution and the information are made available to them. Under what conditions do they make the effort of thoroughly checking the quality of the alternative solution? When do they choose to ignore it and simply stay with the current solution? Above all, we are interested in the information acquisition strategies that precede Adapt/Exchange decisions. Therefore, the following section will discuss what we can learn about such information acquisition strategies from research in other decision contexts.

**Figure 1**

Implementation of the abstraction hierarchy in the present study. When adapting the current solution (e.g., turnover wheel) by modifying its properties (e.g., material) does not meet the requirements, problem solvers need to move up in the hierarchy and reconsider the solution principle as such, in this case by exchanging the type of component (e.g., using a turnover belt instead).

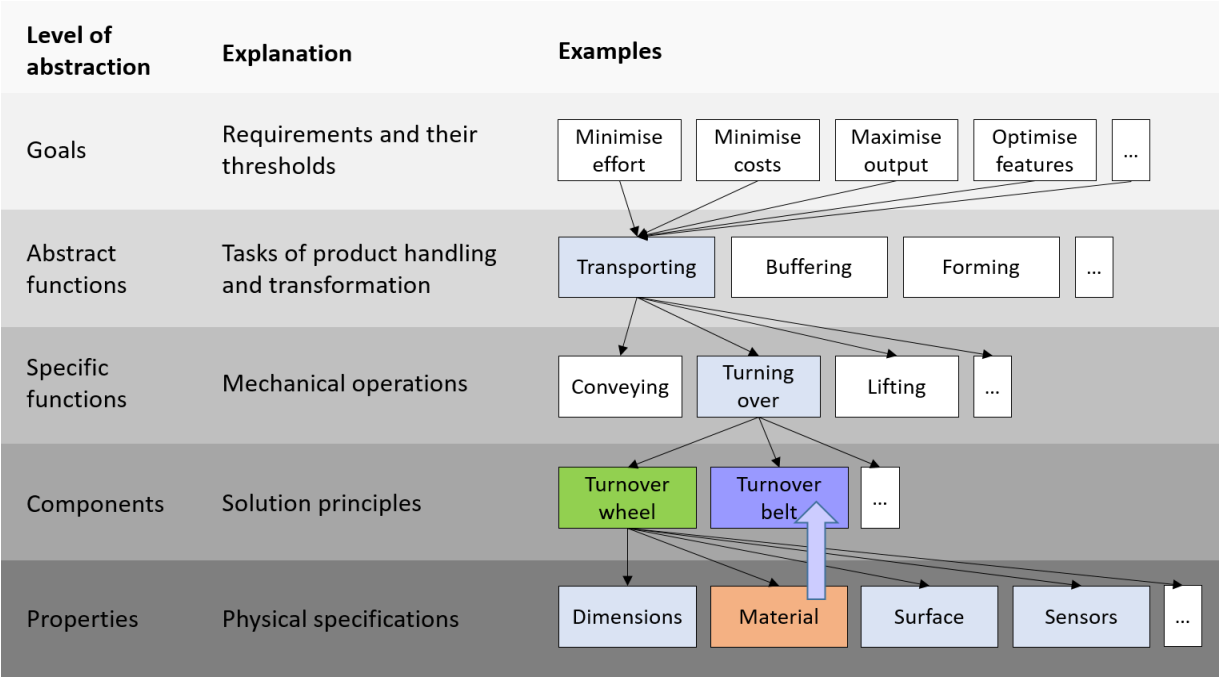

### 1.3 How do decision makers acquire information about potential solutions?

Several lines of decision making research have investigated how people collect and integrate information before making their choices. In this section, we will consider some of them, and compare their problem structures to that of Adapt/Exchange decisions. As we will see, these problem structures differ quite starkly, limiting the inferences we can draw from previous research, and emphasising the need for dedicated studies of information acquisition in Adapt/Exchange decisions.

In *multi-attribute decision making*, people choose between several options that differ in several attributes. For instance, they might choose between different houses that differ in their location, price, presence of a garden, and connection to public transport. People typically do not consider all attributes and integrate them in a weighted-average manner, but instead use simpler heuristics (Bettman et al., 1993; Payne et al., 1988). This is the case especially when interpreting or integrating the information is effortful, for instance when the scales and categories are ill-designed or when the attributes must be sampled successively (Glöckner & Betsch, 2008; Söllner et al., 2013). Similarly, people show various distortions and biases in the way they use particular attributes (Martin & Norton, 2009; Sun et al., 2010). These findings cast doubts on whether people will thoroughly check all available information in Adapt/Exchange decisions. However, note that in multi-attribute decision making, limited information acquisition usually pertains to attributes rather than options – people rarely ignore an option as a whole. Thus, the problem structure differs from Adapt/Exchange decisions, where we assume that people might refrain from checking one of only two solutions altogether. Moreover, given that they do check the Exchange solution, we assume that they will most likely examine all of its attributes (e.g., check whether a component meets all requirements, not just an arbitrary subset). Therefore, the conclusions we can draw from multi-attribute decision making research are limited.

Another line of research has investigated the conditions under which people stop their information acquisition efforts once an option is good enough. This is referred to as *satisficing* (Simon, 1956, 1990), and it typically occurs when the number of options is excessive or unknown (Gigerenzer et al., 2012; Todd & Gigerenzer, 2007; Todd & Miller, 1999). For instance, when choosing a spouse, people cannot possibly consider all options that may exist in principle (Todd & Miller, 1999). Although satisficing is considered a heuristic, it can be a highly reasonable strategy in complex environments (Todd et al., 2012). However, this line of research also does not enable strong inferences about information acquisition strategies in Adapt/Exchange decisions. This is because it is unclear whether satisficing also occurs when the number of options is as low as in Adapt/Exchange decisions, where the choice is often restricted to only two solutions (Müller & Urbas, 2017, 2020). Can satisficing be an issue in this case?

It might be, namely when the *costs of information acquisition* are high. For a long time, it has been known that the higher the costs of evaluating an option, the fewer options are considered (Hauser & Wernerfelt, 1990). However, not until recently, studies have begun to systematically examine how decision makers trade off the costs and benefits of information acquisition (Fiedler et al., 2021; McCaughey et al., 2023). For instance, in a series of experiments (Fiedler et al., 2021), participants could sample information about the past performance of two stocks, and each piece of information reflected an exemplary outcome (i.e., increase or decrease). Thus, the more often participants sampled the stocks, the more accurately they could assess which one is superior – but the longer it took them to reach a decision and thus the fewer decisions they were able to make in the available time. When do people stop gathering information and make their choice? The authors observed a striking tendency to over-sample. Participants did not seem to consider that collecting more samples had a less beneficial impact on accuracy than it had a detrimental impact on speed, and thus produced much lower payoffs overall. This is interesting, given that previous research has often stressed that people fail to acquire sufficient information to make optimal choices. However, when information does not come for free, the principle of "more information is better" does not hold anymore, and the optimal amount depends on the costs of information acquisition (McCaughey et al., 2023). In spite of this, participants in psychological studies usually prioritise accuracy over speed and efficiency (Fiedler & McCaughey, 2023). Thus, in Adapt/Exchange decisions, people might also show an accuracy bias, consistently checking the alternative solution even when the current one is feasible and information acquisition is costly. However, two factors set Adapt/Exchange decisions apart from the information sampling studies discussed above. The first concerns the type of information acquisition. In the sampling studies, participants checked how often an option produced a particular outcome, while in our Adapt/Exchange context, they check how well a solution can solve the problem right now. The former requires a quantitative integration of information across instances, while the latter requires a qualitative integration within a single instance. It is unclear how this difference will affect information acquisition strategies. A second difference between sampling studies and Adapt/Exchange decisions concerns the dominance of options. In the sampling studies, participants encountered two parallel options with equal weight, while in Adapt/Exchange decisions, one solution (i.e., Adapt) represents the status quo. How might information acquisition be affected by non-equally weighted solutions?

Information acquisition in the presence of recommended solutions has been the focus of Human Factors studies on the use and misuse of *decision support systems*. When such systems offer a solution, do people directly accept it, or do they carefully cross-check its validity? The evidence is mixed. On the one hand, a myriad of studies has reported that people over-rely on automated systems and fail to cross-check their decisions – a phenomenon known as *automation bias* (for reviews see Mosier & Manzey, 2019; Onnasch et al., 2014; Parasuraman & Manzey, 2010). Automation bias might increase with verification complexity (i.e., costs of information acquisition), but the evidence is limited so far (Lyell & Coiera, 2017). On the other hand, under some circumstances people do cross-check suggested solutions quite thoroughly, even when information acquisition is costly. For instance, when people had to diagnose and solve machine problems and a particular solution was highly recommended, they thoroughly cross-checked its validity, even when time pressure and workload were high (Müller et al., 2019). Similarly, people usually cross-checked the validity of alarms, even when the time-consuming checks drew resources from concurrent tasks (Manzey et al., 2014). The task features that determine when people cross-check recommended solutions are yet to be clarified (for a discussion see Müller et al., 2019). Still, in line with the over-sampling studies discussed above, a take-home message is that people sometimes have a strong preference for accuracy, despite substantial costs of information acquisition. This might lead to the conclusion that that they will also consistently check the alternative solution in Adapt/Exchange decisions.

Taken together, previous research on information acquisition during decision making has painted a mixed picture. On the one hand, when people choose between multiple options with multiple attributes, they usually do not collect and integrate all available information. This is exacerbated when the information is suboptimally presented and when the available options are excessive or unknown. Can such limited information acquisition also be expected with only two options? One might assume that this depends on the costs of information acquisition. However, according to basic research on information sampling and applied research on using automated decision support, people often incur substantial costs for the sake of making slightly more accurate decisions. Thus, one might infer that they will also scrutinise the alternative solution in Adapt/Exchange decisions, even when the current solution is feasible. However, this contrasts with the speculative conclusion from a previous study (Müller & Urbas, 2020), where performance outcomes suggested that people refrained from checking the alternative Exchange solution as long as the current Adapt solution was good enough. Unfortunately, this conclusion remained highly speculative, because information acquisition was not actually assessed. To close this gap, an experimental setting is needed that allows us to trace the process of how people acquire information while making Adapt/Exchange decisions.

## 1.4 Present study

*1.4.1 Research questions end experimental setup*

The aim of the present study was to broaden our understanding of information acquisition processes in Adapt/Exchange decisions. Specifically, we investigated under what conditions people check the alternative solution. Do they consistently engage in a thorough comparison of both solutions before making their choice, or do they only consider alternative solutions when adaptations of the current solution are unable to solve the problem? In other words, do they refrain from checking Exchange when Adapt is good enough?

We addressed this question in a computer-based mechanical engineering task, in which participants had to select and specify a component for a chocolate wrapping machine. This component fulfilled the function of turning over a chocolate bar, which could be realised by two different solution principles. These solution principles were made available to participants by a simulated decision support system. One solution was framed as a previously used machine component that participants could modify in terms of its specific physical properties (Adapt), whereas they were instructed that the alternative solution (Exchange) had to be constructed from scratch. However, this additional construction process was only hypothetical and not part of the experimental task, which was identical for both solutions. In each trial, participants performed a multi-step optimisation procedure (for an overview see Figure 3). First, they had to select which solution they wanted to check and specify (e.g., adapt the current turnover wheel, or exchange it for a turnover belt). Second, they had to check the feasibility and quality of this solution in four steps of specifying its physical properties (e.g., size of the turning wheel). To this end, they had to check which of two versions of the respective property (e.g., two sizes) performed better on four requirements (i.e., effort, costs, output, physical features). To do this, they had to compare the versions' numerical goal achievement values (e.g., their costs) to a fixed requirement threshold (e.g., maximum acceptable cost). After this process of checking and specifying was completed, participants could either confirm their solution (with all adaptations they had made) or check the other solution principle (e.g., exchange the turnover wheel for a turnover belt).

Our aim was to examine how the frequency of checking Exchange depended on whether Adapt was good enough. Therefore, we manipulated the feasibility of both solutions between trials. The feasibility of Adapt reflected whether the requirements could be met by merely adapting the current component (see Table 1). Thus, participants could only find out whether Adapt was feasible by checking all of its properties. Adapt could either be non-problematic, meaning that it met all requirement thresholds (henceforth referred to as A+), locally problematic in the sense that it failed to meet an individual requirement threshold for one property but still met the total requirement threshold at the end of the trial (A(-)), or globally problematic in the sense that it also failed to meet this total threshold (A-). Local problems were undesirable but acceptable, whereas global problems rendered Adapt impossible. Exchange always met all individual and total requirement thresholds, as this is the very purpose of exchanging the solution principle. Instead, the feasibility of Exchange depended on whether the new component could be constructed before the scheduled delivery deadline (E+) or not (E-). This was already known at the start of a trial and did not require any extra checking. Checking Exchange could still be useful, because the two solutions reached different point scores with regard to the requirements, determining which one was numerically better. However, checking Exchange never was mandatory, because as long as no global problems (A-) or scheduling problems (E-) were present, either solution was a valid choice.

**Table 1**

Five types of solution feasibility with their labels and explanations for Adapt and Exchange. Selecting a solution with a "+" or "(-)" sign was a valid choice, while selecting a solution with a "-" sign was an error

| Label | Adapt | Exchange |
| --- | --- | --- |
| A-E+ | Global problem: a total threshold on one requirement is not met after specifying all properties | No problem: required construction time is shorter than time to deadline |
| A(-)E+ | Local problem: an individual threshold on one requirement is not met for one property | No problem: required construction time is shorter than time to deadline |
| A(-)E- | Local problem: an individual threshold on one requirement is not met for one property | Scheduling problem: required construction time is longer than time to deadline |
| A+E+ | No problem: all thresholds are met | No problem: required construction time is shorter than time to deadline |
| A+E- | No problem: all thresholds are met | Scheduling problem: required construction time is longer than time to deadline |

*1.4.2 Hypotheses*

*Information acquisition.* When do people check the alternative solution in Adapt/Exchange decisions? Below we present four competing hypotheses about the percentage of trials in which participants check Exchange. Note that we generally assumed that participants rarely check Exchange when it is ruled out in advance by a scheduling problem (E-). Similarly, we assumed that participants consistently check Exchange when Adapt is ruled out by a global problem (A-). Accordingly, these conditions served as an upper and lower baseline, respectively.

In principle, one conceivable outcome is what we will call the *no-checking hypothesis*. It rests on the large literature on automation bias, reporting that people uncritically accept the solutions suggested by decision support systems (Mosier & Manzey, 2019; Onnasch et al., 2014; Parasuraman & Manzey, 2010). Similarly, they might accept Adapt without checking its feasibility, and thus also have no need to check Exchange. Regardless of Adapt feasibility, Exchange checking rates should not differ from the lower baseline. That is, A-E+, A(-)E+, and A+E+ should neither differ from A+E-, nor from each other. However, we considered this hypothesis highly unlikely, as the present task structure clearly differed from that of typical studies on automation bias. We will elaborate these differences in the Discussion.

The second hypothesis will be referred to as the *consistent-checking hypothesis*. It rests on the observation that people have a strong bias for accuracy, and thus thoroughly acquire information even when it is costly and the achievable gains are minimal (Fiedler et al., 2021; Manzey et al., 2014; Müller et al., 2020; Müller et al., 2019). Accordingly, participants should be highly motivated to find out which solution reaches a higher point score, and therefore check both solutions unless the alternative is ruled out in advance (E-). Thus, similar to the no-checking hypothesis, the frequency of checking Exchange would not depend on Adapt feasibility, but it should be similar to the upper instead of the lower baseline. That is, A(-)E+, and A+E+ should neither differ from each other, nor from A-E+.

The hypothesis we considered most likely was that people satisfice when making Adapt/Exchange decisions (Müller & Urbas, 2020). This satisficing hypothesis comes in two flavours. The *strong satisficing hypothesis* predicts that participants refrain from checking Exchange whenever Adapt is good enough. Thus, when Adapt is non-problematic, the frequency of checking Exchange should not differ between situations in which Exchange is possible (A+E+) or impossible (A+E-, lower baseline). The *weak satisficing hypothesis* predicts that the rate of checking Exchange decreases as Adapt feasibility increases. Most Exchange checking is expected when Adapt is globally problematic (A-E+), less when Adapt is locally problematic (A(-)E+), and even less when Adapt is non-problematic (A+E+).

We also analysed the frequency of checking Adapt in an exploratory manner. We had no specific hypotheses about that, other than assuming that participants would always check this solution. This is because participants had to go through the specification procedure of at least one solution to reach the end of the trial. We expected this to be Adapt in any case, given that it was the default suggested by the system. Moreover, it also did not require additional construction effort, making it a priori preferable to Exchange.

*Choice.* In addition to participants' information acquisition behaviour, we also analysed their rates of choosing Exchange. Regardless of whether the information acquisition results would support the consistent-checking hypothesis or either version of the satisficing hypothesis, we expected the same data pattern: participants' choices should reflect the different levels of Adapt and Exchange feasibility. That is, we expected most Exchange choices when Adapt was globally problematic (A-), fewer when it was locally problematic (A(-)), and fewest when it was non-problematic (A+). Similarly, we expected more Exchange choices when Exchange was non-problematic (E+) than when it was ruled out by a scheduling problem (E-). Thus, regardless of participants' information acquisition behaviour, we expected their choice patterns to resemble the information acquisition pattern of the weak satisficing hypothesis. Finally, we expected a general increase in Exchange choices when it reached a higher point score than Adapt, but had no hypotheses about interactions of this factor with solution feasibility.

*Exploratory analyses of individual strategies.* We intended to assess inter-individual differences in participants' information acquisition strategies in an exploratory manner. Specifically, we looked at the order in which participants would check the two solutions (i.e., which solution is checked first, and whether solutions are checked repeatedly).

## 2. Methods

### 2.1 Data availability

All stimuli, instructions, human participant data, and syntax files are made available via the Open Science Framework: https://osf.io/53yd4/

### 2.2 Participants

Twenty-five members of the TUD Dresden University of Technology participant pool (ORSEE, Greiner, 2015) took part in the study in exchange for course credit or 8€ per hour. No participant had to be excluded, because all achieved the required score in a knowledge test before the experiment (see below). The sample included 17 female and 8 male participants, and their age ranged from 18 to 45 years ($M$ = 26.4, $SD$ = 6.4). Participants provided written informed consent and all procedures followed the principles of the Declaration of Helsinki.

## 2.3 Apparatus and stimuli

*2.3.1 Lab setup*

Experiments took place in a quiet lab room, using one of two desktop computers (monitor sizes 24") for stimulus presentation as well as a standard QWERTZ keyboard and computer mouse as input devices. The experimental procedure was programmed with the Experiment Builder (SR Research, Ontario, Canada, Version 2.2.61).

*2.3.2 Instruction video and knowledge test*

Before starting the experiment, participants watched an instruction video of 22 minutes. The video was based on a Microsoft PowerPoint presentation, which used several instructional techniques to facilitate learning, such as animations (Mayer & Moreno, 2003), advance organisers (Mayer, 2008), test questions (Nungester & Duchastel, 1982), and summary slides. The video consisted of two parts:

*Introduction of the scenario*. Participants were informed that during the experiment they would play the role of an engineer designing a component of a wrapping machine for chocolate bars. They learned that the function of this component was to turn a chocolate bar, and that two solution principles were available. The video explained that each solution could be specified by selecting between two versions for each of four properties. Participants learned that each version was characterised by its values with regard to four requirements, and that it was important to achieve low values on the first two requirements, but high values on the second two requirements.

*Experimental screens and procedure*. The second part of the video provided walk through a step-by-step demonstration of the experimental procedure. It explained how the solutions had to be selected and specified, by presenting the consecutive screens of an example trial. The contents of each screen were described, and participants were shown which actions they could perform. However, no strategy for comparing the values or making the decisions was suggested. In the example, Adapt was selected, specified, and confirmed. Participants were informed that they could go back and check out Exchange as well (following the same procedure), but this procedure was not shown to them again.

*Knowledge test*. A paper-and-pencil knowledge test was administered to identify participants with insufficient understanding of the instruction. This test consisted of 10 multiple-choice questions with four response alternatives, only one of which was correct for each question. A cut-off of 90 % correct answers was used to determine whether a participant needed to be excluded from the experiment.

*2.3.3 Scenarios and calculations*

The scenario reflected one part of mechanical engineering processes: the selection and specification of solution principles (for an overview see Table 2). The solutions were two machine components, a turnover wheel and a turnover belt. They could be specified regarding four properties (i.e., dimensions, material, surface, and sensors) by selecting among two versions (e.g., 5 vs. 7 cm for dimensions). The two versions differed in their performance on four requirements (i.e., effort, costs, output, physical features). All properties and requirements were treated alike in the calculations (see below). Thus, in principle participants could simply ignore their identities (e.g., whether a property was dimensions vs. material, or whether a requirements was effort vs. costs) and simply add up their numerical values.

*Thresholds*. The individual thresholds for the property/requirement combinations had values of either 40, 50, 60 or 70, which were counterbalanced across the four properties (see lower part of Table 2). Accordingly, summed over all properties, the total threshold for each requirement was 220. The individual and total thresholds needed to be considered when specifying a solution. For the requirements effort and costs, the values of a selected version had to be lower than the thresholds, and for output and physical features they had to be higher. The actual requirement values for the two versions were created by generating random numbers: If the value had to be lower than the threshold (for effort and costs), random values between one and ten points below the threshold were generated (e.g., between 60 and 69, if the threshold was < 70). If the value had to be higher than the threshold (for output and physical features), random values between one and ten points above the threshold were generated (e.g., between 71 and 80, if the threshold was > 70).

*Solution feasibility*. The feasibility of Adapt and Exchange solutions varied between five types of trials (see Table 1). For Adapt, solution feasibility was defined by whether it was possible to meet the requirement thresholds. Adapt could either be non-problematic, locally problematic, or globally problematic. (1) Non-problematic (A+) meant that all eight requirement values (i.e., four properties with two versions each) met the individual thresholds. (2) Locally problematic (A(-)) meant that one requirement threshold for one property could not be met by any of the two versions (e.g., in the property "dimensions" neither the 5 nor 7 cm version met the effort threshold of < 70). The threshold was violated by 11 to 19 points. However, this did not lead to a global problem, as it was compensated by the remaining properties. (3) Globally problematic (A-) meant that in addition to a local problem, the total achieved value on one requirement did not meet the acceptable total threshold of 220 at the end of the property specification, no matter which versions were chosen during the trial. The total achieved values were calculated by adding the values of the four chosen versions in the four properties. The total threshold of 220 was at least violated by one to three points, if participants had chosen optimally with regard to the respective requirement, but this deviation could be higher, depending on participants' choices. Local problems led to global problems in 50 % of the trials, whereas all trials with a global problem also had a local problem.

Contrary to Adapt, Exchange always met the individual and total requirement thresholds. Instead, its feasibility was defined by whether the deadline for delivering the machine was shorter or longer than the time needed to construct the new component. Both the delivery deadline and the required time were presented as numbers of days. (1) No problem (E+) meant that more days were left until the delivery deadline than days needed for construction, whereas (2) a scheduling problem (E-) meant that fewer days were left. For the days left until the delivery deadline, random values between 21 and 30 were generated. The difference between the delivery deadline and the required time was created by selecting random values between 5 and 9. Based on this, the days at least required for the construction were calculated: If Exchange had no problem, the difference was subtracted from the days left until the delivery deadline (e.g., 25-7=18). If Exchange had a scheduling problem, the difference was added to the days left until the delivery deadline (e.g., 25+7=32).

All combinations of Adapt and Exchange feasibility were presented in the experiment, except for a global problem combined with a scheduling problem (A-E-), because such trials would be non-solvable.

*Solution with higher point score*. The trials differed as to whether Adapt or Exchange reached a higher point score at the end of the property specification procedure, if always choosing optimally (i.e., choosing the versions with the higher deviation from the threshold in the desired direction). This point score was created by adding the amounts by which each total requirement value differed from its total threshold. For instance, when the total requirement values across the four properties were 191 for effort, 195 for costs, 236 for output and 240 for physical features, the point score corresponded to the sum of their differences from 220, which is 93. The differences between the point scores achievable with Adapt versus Exchange ranged from 7 to 32. Importantly, which solution reached a higher point score was independent of solution feasibility. For instance, even when Adapt was globally problematic, it could still reach a higher point score than Exchange, given that its violated threshold on one requirement was compensated by high values on the other requirements. This enables a factorial combination of solution feasibility and solution with higher point score.

**Table 2**

Overview of the scenario. Each solution had four properties, each property had two versions, and the thresholds of four requirements had to be met during each step of solution specification

|  | Property 1: Dimensions | Property 2: Material | Property 3: Surface | Property 4: Sensors |
|---|---|---|---|---|
| Solution principle: Turnover wheel | | | | |
| Version 1 | 5 cm | Steel | Coated | Without |
| Version 2 | 7 cm | Aluminium | Non-coated | With |
| Solution principle: Turnover belt | | | | |
| Version 1 | 13 cm | Rubber | Grooves | Without |
| Version 2 | 17 cm | Fabric | No grooves | With |
| Requirements | | | | |
| Effort | < 70 | < 40 | < 60 | < 50 |
| Costs | < 40 | < 70 | < 50 | < 60 |
| Output | > 50 | > 60 | > 40 | > 70 |
| Physical features | > 60 | > 50 | > 70 | > 40 |

*2.3.4 Screens*

Stimuli were shown with a resolution of 1920 × 1080 pixels. All stimuli presented text, pictures, and interaction elements on a black background and with consistent colour coding (i.e., green for Adapt, purple for Exchange). All text was presented in German. Example stimuli are shown in Figure 2. Each trial used three types of screens: a principle selection screen as well as four choice screens and one summary screen for each solution principle. Additionally, a number of instruction screens were shown.

*Principle selection screen.* The principle selection screen (see Figure 2A) presented the task, two boxes for the solutions, and two buttons to check these solutions. The task shown at the top of the screen specified the intended machine function (i.e., "Turn over a chocolate bar!") and remained the same throughout the experiment. The two boxes always put Adapt on the left-hand side and Exchange on the right-hand side. They contained a description of the respective component, its name and a picture, as well as additional information about it. The Adapt box stated that a solution already existed (e.g., turnover wheel) and that participants could adopt or partly adapt it. In addition to the solution's name and picture, its current properties were listed (e.g., dimensions: 5cm). The Exchange box stated that alternatively, another solution principle could be applied, but that it would have to be designed from scratch. Moreover, in addition to the name and picture, two temporal information elements were provided: a delivery deadline indicating how many days were left until the machine had to be delivered to the customer, and an estimation of the minimum number of days that the construction process would take when exchanging the component. Below each box, a button allowed participants to check the current or alternative solution (i.e., Adapt and Exchange, respectively).

*Choice screens*. The choice screens (see Figure 2B and C) provided information about the specification of each property of the respective solution. The upper part of the screen presented the solution (e.g., Adapt), the name of the component (e.g., turnover wheel), and a photograph of the implemented solution. The middle part of the screen specified the current properties (e.g., material), current values, and thresholds. For the current property, two versions were available (e.g., rubber and steel). Adapt and Exchange slightly differed in how these versions were presented. For Adapt, version 1, which corresponded to the current implementation, was labeled "existing solution" and highlighted by a blue box, while version 2 was labeled "Adapt" and presented without highlighting. For Exchange, the two versions were labeled "option 1" and "option 2", respectively, and none was highlighted. For each version, a table listed its values on four requirements (i.e., effort, costs, output, physical features). The current values provided the sum for each requirement based on the previous choices. The thresholds indicated the maximum value (for effort and cost) or minimum value (for output and physical features) that needed to be achieved. These thresholds were provided for both the current trial and the total (i.e., sum over all four properties). As the current thresholds *should* be complied with, they were highlighted in yellow, and as the total thresholds *had to* be complied with, they were highlighted in red. The lower part of the screen provided buttons for the available actions. Below each version, a button allowed participants to select it. For Adapt, version 1 corresponded to the current solution and version 2 corresponded to an adaptation of it. Hence, the buttons were labeled "Adopt" and "Adapt", respectively. For Exchange, both versions corresponded to a new design, and hence both buttons were labeled "Select". Finally, a "To Principle Selection" button in the lower right corner allowed participants to return to the principle selection screen.

*Summary screen*. For each solution principle, a screen summarised the specifications of the four properties that had been selected on the previous choice screens (see Figure 2D). The upper part of the summary screen was identical to the choice screens, naming and visualising the solution. The middle part provided a list of the selected property versions (e.g., dimensions: 5 cm, material: steel, surface: coated, sensors: no), the sum of the achieved values for each requirement, and the total thresholds highlighted in red. Moreover, it provided a total point score indicating the summed deviations of the achieved values from the thresholds in the desired direction (e.g., a value of 93 could indicate that effort deviated by 29, cost by 25, output by 16, and physical features by 20). The lower part of the screen provided a button to confirm the current solution and a button to return to the principle selection screen.

*Additional instruction screens*. On an opening screen, participants were welcomed, informed that demographic data would be requested next, and that they could proceed with the space bar. On a demographics screen, participants had to fill in their age and gender. Two introductory screens informed participants that the experiment would start, and that a practice trial together with the experimenter would follow. After practice, a start screen stated that the main experiment would begin and that participants could take breaks between trials. After each trial, intermediate screens announced the next trial. Finally, an ending screen told participants that the experiment was finished.

**Figure 2**

Stimulus material. (A) Principle selection screen. (B) Choice screen Adapt. (C) Choice screen Exchange. (D) Summary screen Adapt. In the experiment, all text was presented in German.

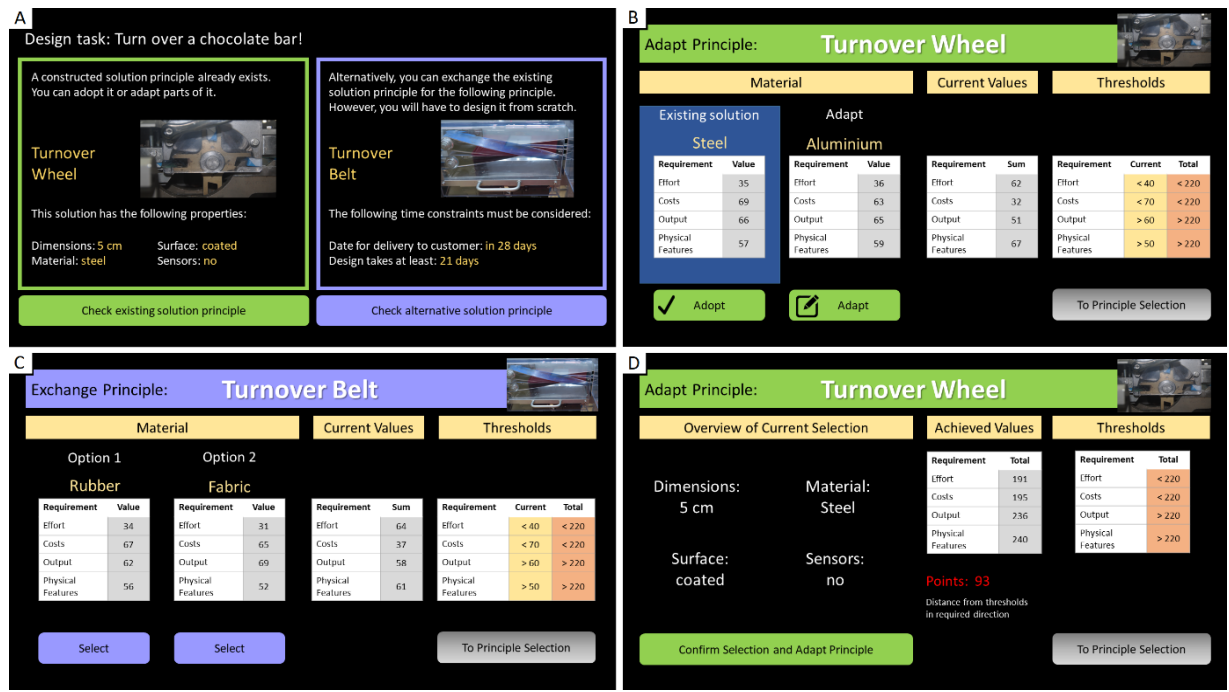

## 2.4 Procedure

*2.4.1 Instruction video and knowledge test*

Participants watched the instruction video and then completed the multiple-choice test. For being allowed to participate in the experiment, they had to answer at least 90 % of the questions correctly. All participants passed the test. The instruction video and knowledge test took about 30 minutes.

*2.4.2 Experiment*

The experiment consisted of one practice trial and thirty experimental trials. The practice trial was an A+E+ trial, meaning that there were no local or global Adapt problems and no Exchange scheduling problems. Adapt reached the higher point score. Participants went through this practice trial with the experimenter and could ask questions. Afterwards, they completed three blocks of ten experimental trials. These ten trials corresponded to all combinations of the two within-participants factors *solution feasibility* (A-E+, A(-)E+, A(-)E-, A+E+, A+E-) and *solution with higher point score* (Adapt, Exchange). The order of trials in a block was randomised for each participant individually, and the order of blocks was counterbalanced across participants. It also was counterbalanced which component (i.e., turnover wheel, turnover belt) served as the current (Adapt) or alternative (Exchange) solution.

Participants could already infer the feasibility of Exchange on the principle selection screen by comparing the available time to the delivery deadline. To infer the feasibility of Adapt, they had to go through all four choice screens to check the four properties (i.e., calculating whether the solution met the requirement thresholds). In A- and A(-) trials, there was one property (e.g., dimensions) for which neither of the two versions (e.g., 5 and 7 cm) met one of the requirements (e.g., both were too costly). The property and requirement on which this problem showed up were selected pseudo-randomly, with the constraint of being evenly spread across trials instead of being biased towards a particular requirement or property. In the same way, we pseudo-randomised the requirement on which the total threshold was not met at the end of the Adapt checking procedure in A- trials.

A schematic overview of the trial procedure is provided in Figure 3. Each trial started with the principle selection screen on which participants had to select whether they wanted to check and specify the pre-constructed solution (Adapt) or the alternative solution (Exchange) that needed to be constructed from scratch. If participants wanted to check Adapt, they had to click the left green button ("Check existing solution principle") on the principle selection screen, which led them to the first choice screen for Adapt. If they wanted to check Exchange, they had to click the right purple button ("Check alternative solution principle"), which led them to the first choice screen for Exchange. The procedure of checking and specifying a solution was identical for Adapt and Exchange. Therefore, in the next paragraph we only describe the procedure for Adapt.

**Figure 3**

Procedure of a trial. Participants first selected a solution principle to check, then specified its four properties by choosing between two versions, and finally confirmed the specified solution. During each step, they could return to the principle selection and check the other solution. Adapt could either meet or not meet the individual and total requirement thresholds, while Exchange always met them but could either meet or not meet a delivery deadline. P = property, V = version, R = requirement

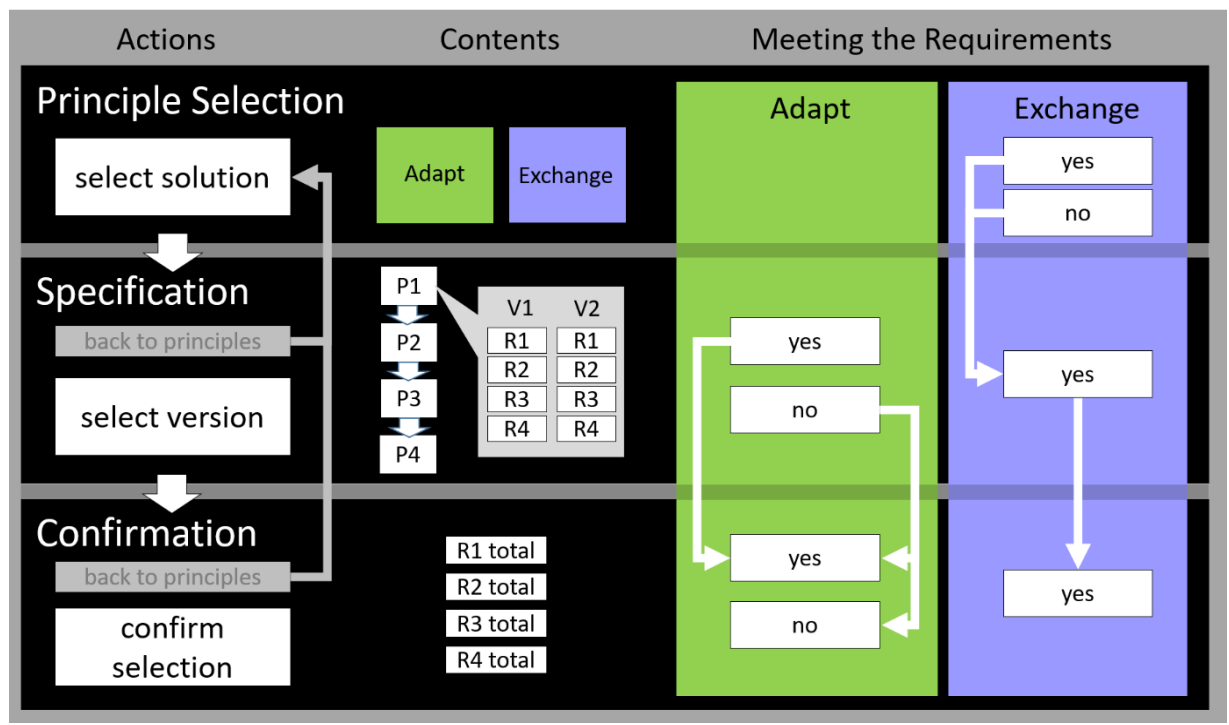

In a five-step procedure, participants selected between two versions for each of the four properties and then entered the summary screen to either confirm their specifications or go back to the principle selection screen. In detail, they first had to specify the dimensions of the component by either adopting the existing property version presented on the left (5 cm), or by adapting it to a different property value presented on the right (7 cm). To this end, participants had to compare the two versions with regard to the four requirements (i.e., effort, cost, output, and physical features), considering both the current and total requirement thresholds. For instance, if the acceptable cost threshold is 70 and this value must not be exceeded, you might prefer version 2 with a cost of 63 to version 1 with a cost of 69, because the former is further away from the threshold in the desired direction. When finished with this comparison, participants could either choose the available property version by clicking the left button ("Adopt"), or the adapted property version by clicking the right button ("Adapt"). Alternatively, they could abort their checking procedure and return to the principle selection screen by clicking the "To Principle Selection" button. Choosing a property version for dimensions brought participants to the second choice screen on which they had to perform the same selection procedure for the property material (i.e., steel vs. aluminium). Subsequently, they were transferred to the third and fourth choice screen for the properties surface and sensors, respectively. After having specified all fourth properties, the summary screen appeared (see Figure 2D) on which participants could compare their achieved requirement values with the total thresholds, and view their achieved point score. They could either confirm their previous actions by clicking the "Confirm Selection and Adapt Principle" button, which brought them to the next trial, or return to the principle selection screen by clicking the "To Principle Selection" button.

After three blocks with ten trials each, the experiment ended and participants entered a post-experimental interview to explain their strategies to the experimenter. Finally, they received their course credit or monetary compensation and were debriefed about the specific goals and hypotheses of the study. Taken together, an experimental session took about one and a half hours.

## 2.5 Data analysis

We analysed participants' choices and their underlying information acquisition behaviour. Choices were analysed by calculating the percentage of trials in which participants chose Exchange. Information acquisition behaviour was analysed as the percentage of trials in which participants checked the Adapt and Exchange solution, respectively. That is, an Exchange checking rate of 100 % means that a participant went through the four choice screens of Exchange in every single trial. Only trials were included in which participants clicked through all four properties before either confirming a solution or returning to the principle selection screen, because only complete checks allowed participants to conclusively evaluate the solution. This led to an exclusion of 46 aborted solution checks over all participants (6.1 % of the total checks). Of these aborted checks, 41.3 % were committed by only two participants, suggesting that aborting a check was not a common strategy. Moreover, in 78.3 % of the aborted checks, participants only opened the first property and then immediately returned to the principle selection screen, suggesting that aborted checks did not usually reflect genuine evaluations of a solution.

To statistically analyse how choice depended on the characteristics of the two solutions, we computed a 5 (*solution feasibility: A-E+, A(-)E+, A(-)E-, A+E+, A+E-*) × 2 (*solution with higher point score: Adapt, Exchange*) repeated measures ANOVA. To analyse information acquisition, we computed two one-way 5 (*solution feasibility: A-E+, A(-)E+, A(-)E-, A+E+, A+E-*) repeated measures ANOVAs, one for checking Adapt and one for checking Exchange. If sphericity was violated, the Greenhouse-Geisser correction was applied and the degrees of freedom were adjusted accordingly. To determine statistical significance, an alpha level of $p < .05$ was used, and all pairwise comparison were performed with Bonferroni correction. Exploratory analyses are described in the respective parts of the Results section.

# 3. Results

## 3.1 Percentage of Exchange choices

Overall, participants chose Exchange in 30.9 % of the trials. The ANOVA revealed significant main effects of solution feasibility, $F(2.8,66.6) = 52.419$, $p < .001$, $\eta_p^2 = .686$, and solution with higher point score, $F(1,24) = 20.178$, $p < .001$, $\eta_p^2 = .457$, as well as an interaction that just reached significance, $F(3.0,70.8) = 2.812$, $p = .046$, $\eta_p^2 = .105$ (see Figure 4A and Table 3). The main effect of solution feasibility indicated that the rate of choosing Exchange depended on problems in both solutions. When Exchange was feasible, the results were in line with the weak satisficing hypothesis: Exchange was mostly chosen when Adapt was globally problematic (A-E+), less when Adapt was only locally problematic (A(-)E+), and even less when Adapt was non-problematic (A+E+) (72.0 vs. 44.0 vs. 26.0 %), all $p$s < .03. Conversely, the data did not support the strong satisficing hypothesis: When Adapt was non-problematic, Exchange was chosen more often when it was feasible (A+E+) than when it was not (A+E-) (26.0 vs. 4.7 %), $p = 026$. The main effect of solution with higher point score indicated that Exchange was chosen more often when it scored higher than Adapt (40.5 vs. 21.3 %). Although the same basic choice pattern was observed regardless of which solution scored higher, the significant interaction can be attributed to stronger effects of solution feasibility when Exchange did. However, the difference between A+E+ and A+E- did not reach significance for either type of solution with higher point score alone, neither for Adapt, $p = .184$, nor for Exchange, $p = .054$. Thus, when Adapt was non-problematic, Exchange choices were rare overall, and the feasibility of Exchange did not make a major difference. Admittedly though, with regard to our strong satisficing hypothesis, this is weak evidence at best. Only because a descriptive difference misses significance, this does not warrant inferring that it does not exist.

## 3.2 Information acquisition

Overall, Exchange was checked less than half as often as Adapt (41.7 vs. 87.7 % of all trials). However, these checking rates strongly depended on solution feasibility (see Figure 4B and Table 3). For *checking Adapt*, there was a significant effect of solution feasibility, $F(4,96) = 10.338$, $p < .001$, $\eta_p^2 = .301$. Pairwise comparisons revealed that the effect merely depended on whether Exchange was feasible or not. When it was feasible, Adapt was checked similarly often when it was globally problematic (A-E+), locally problematic (A(-)E+), and non-problematic (A+E+) (80.7, 80.7 and 80.0 %, respectively), all $p$s > .9. When Exchange was not feasible, Adapt was checked almost all the time, and equally often when it was locally problematic (A(-)E-) and non-problematic (A+E-) (both 98.7 %), $p > .9$. Moreover, for all trial types in which Exchange was feasible, Adapt was checked less often than when Exchange was not feasible, all $p$s < .04.

For *checking Exchange*, there also was a significant effect of solution feasibility, $F(4,96) = 55.789$, $p < .001$, $\eta_p^2 = .699$. The pattern of results mirrored the one observed in the analysis of choice behaviour. In line with the weak satisficing hypothesis, Adapt feasibility modulated the rate of Exchange checking, given that Exchange was feasible. That is, Exchange was checked most often when Adapt was globally problematic (A-E+), less when Adapt was only locally problematic (A(-)E+), and even less when Adapt was non-problematic (A+E+) (88.0 vs. 58.0 vs. 37.3 %), all $p$s < .03. However, contrary to the strong satisficing hypothesis, when Adapt was non-problematic, Exchange was checked more often when it was feasible (A+E+) than when it was not (A+E-) (37.3 vs. 8.0 %), $p = .006$.

**Table 3**

Mean rates of choices and information acquisition (in % of trials) as well as their standard deviations (in parentheses) for choice and information acquisition

|  |  | Solution feasibility | | | | |
| --- | --- | --- | --- | --- | --- | --- |
|  |  | A-E+ | A(-)E+ | A(-)E- | A+E+ | A+E- |
| Choice |  |  |  |  |  |  |
| Solution with higher point score | Adapt | 57.3 (31.2) | 30.7 (28.7) | 1.3 (6.7) | 16.0 (29.1) | 1.3 (6.7) |
|  | Exchange | 86.7 (27.2) | 57.3 (36.7) | 14.7 (23.7) | 36.0 (38.4) | 8.0 (17.4) |
| Information acquisition |  |  |  |  |  |  |
| Checking of solution | Adapt | 80.7 (22.4) | 80.7 (24.9) | 98.7 (27.6) | 80.0 (4.6) | 98.7 (6.7) |
|  | Exchange | 88.0 (22.3) | 58.0 (30.5) | 17.3 (21.2) | 37.3 (35.1) | 8.0 (13.7) |

**Figure 4**

Mean rates of choices and information acquisition (in % of trials), depending on solution feasibility. (A) Choices of Exchange depending on which solution reached the higher point score. (B) Checking of the two solutions. (C) Checking of Exchange for the subsample of participants (n = 14) who started with checking Adapt in at least 80 % of the trials. Error bars represent standard errors of the mean

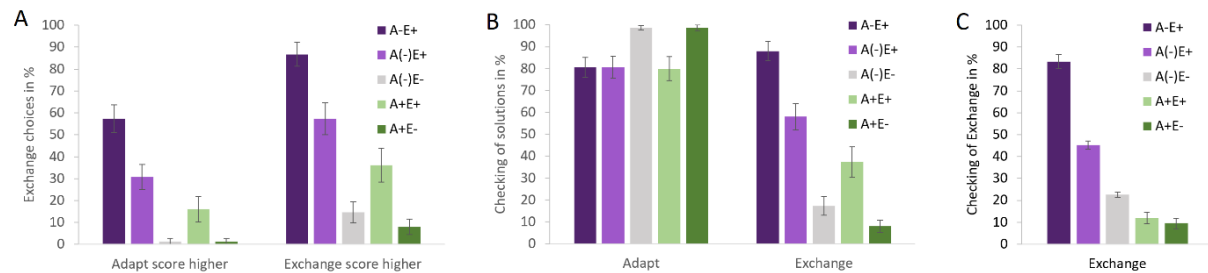

### 3.3 Exploratory analyses

To better understand participants' individual strategies of information acquisition, we plotted which solution they checked at what time of the trial (see Figure 5). An inspection of this figure reveals that, contrary to our assumptions, not all participants started a trial with checking Adapt. In fact, only seven participants always started with Adapt (top row of Figure 5). Other participants started with checking Exchange more or less often. For instance, participants 20-25 (bottom row of Figure 5) preferably started with Exchange when Exchange was feasible (E+). It needs to be noted that both versions of the satisficing hypothesis rested on the assumption that people start with checking Adapt, because this is a necessary condition for knowing whether Adapt is good enough. Therefore, our lack of support for the strong satisficing hypothesis might be attributable to the fact that some participants did not fulfil this basic condition.

**Figure 5**

Individual checking behaviour for all participants and trials. Each black bar represents one participant (sorted by the frequency of checking Adapt first) and the green and purple cells indicate instances of checking Adapt and Exchange, respectively (only complete checks of all four properties are included). The vertical direction represents the 30 trials, sorted by solution feasibility. The horizontal direction represents at what time a solution was checked (e.g., first, second, third). When only green cells are attached to the right of a black bar (i.e., participants 1-7), this means that Adapt was always checked first. When chains of cells extend far to the right, it means that a participant performed several rounds of checking, either by checking the same solution repeatedly (e.g., participant 21, fourth trial of A+E-) or by frequently switching between Adapt and Exchange (e.g., participant 7, last trial of A-E+). The last solution checked in each row is the one ultimately chosen.

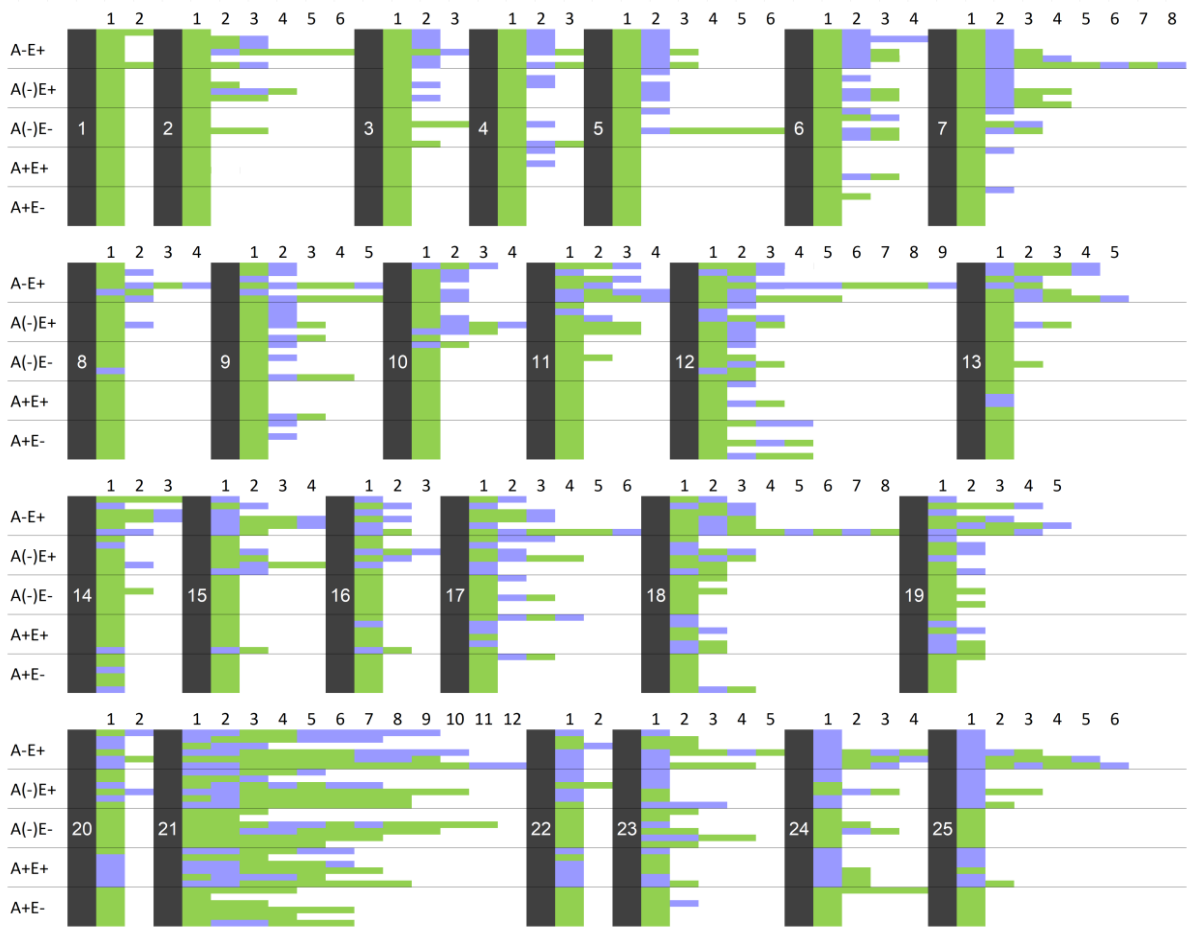

To test this possibility, we repeated the critical information acquisition analysis on the rate of checking Exchange with only those participants who usually fulfilled the necessary condition for satisficing. That is, we computed a one-way 5 (*solution feasibility: A-E+, A(-)E+, A(-)E-, A+E+, A+E-*) repeated measures ANOVA that included only participants who started with checking Adapt in more than 80 % of the trials (including 14 participants, 56 % of the sample). There was a significant effect of solution feasibility, $F(4,52) = 42.278$, $p < .001$, $\eta_p^2 = .765$. In line with the strong satisficing hypothesis, when Adapt was non-problematic, the frequency of checking Exchange did not differ between trials in which Exchange was feasible (A+E+) or not feasible (A+E-) (11.9 and 9.5 %), $p > .9$ (see Figure 4C).

Arguably, the cut-off of 80 % trials that started with checking Adapt is quite arbitrary. Therefore, we repeated the analysis with different cut-offs to test the stability of the strong satisficing hypothesis.

We used cut-offs of 70 % (including 18 participants, 72 %), 50 % (including 21 participants, 84 %) and a median split of the sample (including 13 participants, 52 %). The strong satisficing hypothesis was supported consistently: in all these analyses, the difference between A+E+ and A+E- failed to reach significance, all *p*s > .1. This was true even for the most inclusive cut-off of 50 % trials being started with Adapt (only excluding four participants), *p* = .107. These results indicate that the majority of participants acted according to the strong satisficing hypothesis. This hypothesis only was at odds with the behaviour of a few participants who did not fulfil the necessary condition for satisficing, because they did not find out whether Adapt was good enough in the first place.

# 4. Discussion

Various decision making tasks call for a situation-specific balance between stability and flexibility (Chrysikou et al., 2014; Goschke, 2013; Hommel, 2015). This stability-flexibility trade-off also is at the heart of Adapt/Exchange decisions (Müller & Urbas, 2017). Is it advisable in a given situation to slightly adapt a currently existing solution that has proven successful in the past? Or would it be better to exchange it for a new, less familiar, but potentially superior solution? In an idealised world without resource limitations on the side of the decision maker, the answer should depend on which solution is better able to meet relevant requirements. Thus, decision makers might be expected to carefully evaluate both solutions, and then choose the superior one. However, when balancing stability and flexibility in real life, decision makers face a second trade-off, namely between thoroughness and efficiency (Hollnagel, 2009). This trade-off does not only depend on the solutions as such, but also on the benefits and costs of acquiring information about them (Fiedler & McCaughey, 2023; McCaughey et al., 2023). Is it even worth contemplating alternative solutions? Should decision makers spend their limited resources on checking them out, when the current solution already is good enough? The present study asked how people navigate these trade-offs: under what conditions do they make the extra effort of acquiring information about an alternative solution?

## 4.1 Overview of the study and findings

Participants performed a simplified mechanical engineering task, in which they could either adapt the physical properties of an existing machine component, or exchange this component for a new one. A cognitively demanding multi-step procedure allowed them to check how one or both solutions fulfilled a set of requirements. We varied the feasibility of both solutions to test whether participants would thoroughly check them under all conditions in order to maximise accuracy, or rather show some form of satisficing. A strong version of this satisficing hypothesis holds that participants refrain from checking the alternative Exchange solution altogether, whenever the current Adapt solution can be applied without problems. In this case, Exchange should not be checked any more often when it is feasible than when it is not (i.e., regardless of whether the construction can or cannot be completed in time). A weaker version of the satisficing hypothesis relies on a monotonic dependence of Exchange checking rates on Adapt feasibility. That is, Exchange should be checked less often as Adapt becomes more feasible (from globally problematic to locally problematic to non-problematic).

Our results clearly refuted the hypothesis that participants consistently check both solutions. Even when excluding all trials in which Exchange was known to be non-feasible, it still was only checked in 61.1 % of the trials. Obviously, participants did not attempt to maximise accuracy at any cost. Did they satisfice instead? While our results supported the weak satisficing hypothesis, they contradicted the strong satisficing hypothesis: Exchange checking rates decreased as Adapt became more feasible, but even when Adapt was completely non-problematic, they still exceeded the lower baseline of non-feasible Exchange. However, exploratory analyses of individual participants' information acquisition strategies revealed a crucial dependency: the higher checking rate of feasible than non-feasible Exchange solutions (given that Adapt was non-problematic) could be attributed to only a subset of participants who regularly started their trials with checking Exchange. When excluding these participants, Exchange checking rates no longer depended on Exchange feasibility, as predicted by the strong satisficing hypothesis. Thus, strong satisficing was evident if and only if its necessary condition was met: that people had the chance to initially find out whether Adapt was good enough.

This additional analysis revealed that we had based our hypotheses on an unwarranted assumption. We had pre-supposed a particular strategy, namely that participants would first check Adapt, and then, depending on the outcome, decide whether or not to check Exchange. In reality, however, participants adopted a wide variety of strategies (see Figure 5). Upon closer consideration, the strategy to start with Exchange is quite reasonable. In this way, you only have to go through the cognitively demanding Adapt procedure once, instead of a first time before checking Exchange and then a second time afterwards. However, sometimes participants also started with checking Exchange and then immediately chose it, without checking Adapt at all (in fact, 68 % of the participants did this at least once). Seemingly, satisficing can go both ways – the high information acquisition costs might generally discourage people from checking another solution once they have found one that is feasible and has a reasonably high point score (we will return to this issue below). Overall, these observations are a reminder to be cautious about the implicit assumptions one makes about participants' strategies when formulating hypotheses, as the reality usually is more complex than anticipated. The following sections will discuss our findings in the light of previous research, highlight some limitations of the present study, and propose implications for future work.

## 4.2 Why did participants not check both solutions consistently?

An alternative hypothesis that we contrasted with satisficing was that participants consistently check both solutions. This hypothesis was based on the strong accuracy bias reported in previous studies across various decision contexts (Fiedler & McCaughey, 2023; Fiedler et al., 2021; Manzey et al., 2014; Müller et al., 2020; Müller et al., 2019). However, we found no evidence for consistent (over-)checking. What features of our Adapt/Exchange decisions might have prevented such thorough performance? We can conceive of two potential explanations, which are not mutually exclusive.

First, the *cognitive demands of checking* a solution were considerably higher than in previous studies. Participants had to perform mental arithmetic in a multi-step procedure, and keep track of the results throughout the entire trial. If they decided to check another solution, this meant that they had to memorise the point score of the previously checked solution in order to later compare both scores. This sets our scenario apart not only from simplistic decision making tasks (e.g., Fiedler et al., 2021), but also from the Human Factors studies in which people incurred high information acquisition costs in order to make accurate decisions (e.g., Manzey et al., 2014; Müller et al., 2020; Müller et al., 2019). Those checking procedures typically required sequences of manual actions, comparisons of parameter readings to nominal values, and passive waiting. Thus, information acquisition was time-consuming and tedious, but not cognitively demanding, as it was in the Adapt/Exchange scenario. In the future, it will be interesting to systematically investigate whether and how the checking of alternative solutions varies with the type of information acquisition costs. This might also depend on personality constructs like Cognitive Effort Investment (Kührt et al., 2023).

A second reason why participants might not have consistently checked Exchange is the *absence of clear performance criteria*. In previous studies, failing to check a suggested solution could result in making wrong diagnoses (Müller et al., 2020; Müller et al., 2019) or missing valid alarms (Manzey et al., 2014). In the present study, the only wrong choices were those that selected a non-feasible solution (i.e., one that violated a total requirement or could not be delivered in time). Other than that, participants were free to set their own standards. For instance, some participants might have decided that the additional construction effort of Exchange generally is unjustified. This would obviously make it superfluous to check Exchange, unless Adapt was completely non-feasible. If the goal of sampling possible solutions is to find the one that maximises expected utility (Brockbank et al., 2023), people need not sample solutions that they evaluate negatively right from the start. This freedom of evaluation emerges from the fact that different cost metrics of Adapt/Exchange decisions are incommensurable. For instance, does a point score benefit of 15 justify 23 days of constructing the component? Adapt and Exchange are qualitatively different options, characterised by partly non-overlapping attributes. Although this renders decision making less straightforward, it is an important characteristic of many real-world decisions. Still, it has been ignored almost entirely in psychological research, and thus it remains an open question for future studies how people balance qualitatively different costs and benefits.

## 4.3 What can we learn about decision making in Adapt/Exchange scenarios?

Our results corroborate the conclusions from a previous Adapt/Exchange study, which speculated that participants might satisfice, instead of thoroughly checking both solutions (Müller & Urbas, 2020). This had been inferred from two types of performance data. First, the frequency of choosing Exchange had varied with the costs of Adapt, instead of the cost ratio between Adapt and Exchange. Second, solution times had been higher when participants chose Exchange. While we cannot cross-check the former finding in the present study, we replicated the latter: Exchange choices were much slower than Adapt choices (110 vs. 76 sec, respectively), $t(23) = -4.718$, $p < .001$, $d = -.963$. We do not want to put too much emphasis on this finding, because the trial numbers entering the analysis for Exchange choices were quite low and variable, reflecting that some participants rarely chose the alternative solution. Fortunately, the rates of checking Exchange provided a much more direct and informative measure of information acquisition. They fully supported the speculative conclusion from the previous study, namely that participants tend to avoid checking Exchange when Adapt is good enough.

Given the consistent results across different Adapt/Exchange scenarios, a critical question is what insights they provide. Do they teach us anything new, beyond what is already known from traditional decision making research? One novel aspect of the present results is that we found evidence for satisficing with only two options. To recapitulate, previous studies found that people resort to simple heuristics when comparing multiple options with multiple attributes (Bettman et al., 1993; Payne et al., 1988) and that they satisfice when the number of options is large or unknown (Simon, 1956; Todd et al., 2012; Todd & Miller, 1999). Here, we obtained evidence for satisficing with only two options, showing that participants refrained from checking an option entirely under particular circumstances. As discussed above, we attribute this to a combination of participants setting their own subjective performance standards and shying away from the high information acquisition costs. This resonates with previous findings on the consequences of manipulating information acquisition costs. Specifically, the typical heuristics and shortcuts in multi-attribute decision making are observed when participants have to gather bits and pieces of information sequentially, but disappear when all the information is available at once, or when search demands are low (Glöckner & Betsch, 2008; Söllner et al., 2013). More generally, people's use of information sources depends on their accessibility (O'Reilly, 1982).

Conversely, people include fewer options in their consideration sets when evaluation costs are high (Hauser & Wernerfelt, 1990).

One might argue that it is trivial that people rely on less information when acquiring it is costly. However, this raises serious concerns about the generalisability of psychological findings. In most studies of decision making, all information was readily available or could be obtained with minimal effort. How does this relate to real world decisions, where information has to be actively collected, for instance by conducting open-ended internet searches, performing series of measurements, or even hiring other people to provide the information? Is it warranted to isolate the "core processes" of decision making from the tedious processes of getting to know one's options? The reductive approach of psychological research sure has its benefits. However, we argue for a complementary, more ecologically inspired perspective that factors in the costs of information acquisition. Given that people show a tendency to economise during complex problem solving (Schoppek, 2023), realistic information acquisition costs do not seem conducive to a flexible consideration of new solution alternatives.

That said, the propensity to adapt or exchange will also depend on the work domain. Initial evidence in the context of fault diagnosis comes from an investigation of how domain characteristics shape cognitive requirements and diagnostic strategies (Schmidt & Müller, 2023). It was found that malfunctioning components are more often adapted when solving problems of packaging machines, and more often exchanged in the case of cars. Such differences are deeply rooted in the characteristics of different technical domains, instead of merely depending on the human problem solver.

### 4.4 Limitations of the present study

The present study is rooted in a tradition of engineering psychology where research endeavours start from a problem that exits in the real world. We try to extract its problem structure, and transfer it into a scenario that is simple enough to be investigated in the lab, under somewhat controlled conditions. This approach always requires a delicate balance between peeling off enough confounding factors to gain sufficient experimental control, but not peeling off too much of what constitutes the essence of the problem of interest. Finding this balance comes with a number of limitations both in terms of internal and external validity.

Perhaps the main limitation with regard to internal validity is our *number concept*, or more specifically, the absolute values of requirement point scores and time delays. Their variation inevitably led to numerical differences between trials (e.g., Adapt reaching a point score of 93 in one trial and 78 in another). These absolute values were arbitrary and did not systematically vary with our experimental conditions. Moreover, we ignored them in our argumentation, merely basing our hypotheses on solution feasibility (e.g., A+E-). However, this does not mean that our participants also ignored the absolute numbers. Some participants might have decided that Adapt being "good enough" was not only a matter of its feasibility (as we had intended) but of its point score reaching a particular value, or of its performance on their favourite requirement. In the post-experimental interview, some participants indeed reported to have based their decisions about checking Exchange on the absolute point score of Adapt. Others reported having prioritised particular requirements (e.g., considering output and physical features to be more important than effort and costs). Thus, different participants probably set different subjective criteria for satisficing.

A related limitation is that the *point scores depended on participants' choices*, and thus were not comparable between participants. Individual choices could never change the solution feasibility of a trial (e.g., in A- trials, Adapt remained globally problematic no matter what participants chose). Still, the choice-dependence of point scores might have affected participants' information acquisition decisions. That is, one participant might have seen a high point score at the end of checking Adapt, and concluded that it is not worthwhile to check Exchange. In the very same trial, another participant who has made different component choices might have seen a lower point score, and concluded that it is crucial to check Exchange. This results from an important characteristic of Adapt/Exchange decisions, namely that the decision alone does not determine the outcome, as this outcome critically depends on how the decisions are implemented.

The present study also had its share of limitations regrading external validity. One of them is the mismatch between our theoretical conception of Adapt and Exchange as two *fundamentally different solution principles*, and their implementation in a simplistic lab experiment. We had to face constraints of how the engineering solutions can be presented to non-experts in a temporally restricted, artificial experiment with hypothetical decisions. For our participants, the Adapt solutions were just as new as the Exchange solutions, and all differences were merely based on instruction. Also, the deeper implications of either staying on the current level of abstraction (Adapt) or moving up in the abstraction hierarchy (Exchange) could not be appreciated adequately, given that our student participants most likely did not mentally represent the solutions in terms of their functional principles. Finally, participants did not actually have to endure the additional Exchange effort of spending about a month constructing a new machine component. The risks of such simplifications are obvious, and in the future we would like to study how experts make Adapt/Exchange decisions in their actual work practice.

A second limitation of external validity is that the *integration of goal criteria was far too simple*. In reality, Adapt/Exchange decisions require a balance between qualitatively different goals (e.g., effort, costs, output). For instance, does a 5 % increase in output justify a 10 % increase in costs? We also asked our participants to integrate the solutions' performance on different requirements. However, these requirements did not actually have meaning – participants could simply add up the point values. This simple mathematical integration contrasts with the complex decision making required in real life, which should be addressed in future work. For instance, in another Adapt/Exchange study, we assessed how participants integrate the qualitatively different risks of harming the plant and impairing product quality, and how this integration varied with instructions (Müller & Blunk, in preparation).

A final limitation of external validity is the *dependence of information acquisition on short- versus long-term consequences*. In realistic engineering tasks, the information acquisition costs a designer faces right now presumably weigh less than the hassles a customer must endure for the next 30 years after purchasing an inferior machine. One might conclude that our findings will not scale: in reality, people might always make the effort of checking a reasonable alternative solution. However, it should be considered that in reality, checking also is more demanding than in the present study (e.g., performing resource-intensive tests to find out whether a new solution meets the requirements). In addition, uncertainty will be an issue, which we completely eliminated from our experiment. Accordingly, engineers by no means can be expected to always check alternative solutions as a general strategy. However, the exact conditions under which they do this remain to be studied in the future.

### 4.5 Outlook and future research

We are only at the very beginning of understanding Adapt/Exchange decisions, and the opportunities for future research are exciting. A particularly challenging but important issue is to assess how sensitive people are to *the optimum amount of information acquisition* in different situations. Our participants often refrained from checking Exchange – but is this a bad thing? Not necessarily, as suggested by the literature on automation bias and complacency. In this context, frequently cross-checking automated systems is not necessarily optimal, as it draws resources from other important tasks (Manzey et al., 2014; Moray & Inagaki, 2000). How much information is needed to optimise overall performance? The answer depends on the costs of information acquisition and the accuracy that can be gained with more information. Basing one's decisions on only very few observations can be optimal when information acquisition is costly (Vul et al., 2014). In contrast, lower information costs and higher payoffs for accurate decisions call for larger information samples. However, people seem to be quite insensitive to this relation, making surprisingly little adjustments to their information acquisition activities (McCaughey et al., 2023). Examining this sensitivity for Adapt/Exchange decisions requires systematic manipulations of the costs of checking alternative solutions and the performance improvements that these solutions can achieve.

Along the same lines, it would be interesting to know *how thoroughly people sample information* about a solution principle, instead of just whether. This could be investigated in scenarios that provide plenty of information about each solution. Before deciding between Adapt and Exchange, people could sample as much information as they want, and then terminate their search whenever they feel that they have learned enough. The sampling literature provides insights about when people stop sampling (Gonzalez & Aggarwal, 2023; Prager et al., 2023). However, transferring them to Adapt/Exchange decisions is not straightforward. Customers would likely expect designers to consider all relevant requirements when choosing a solution, not just a subjectively selected subset. Limited information sampling still is conceivable, namely with regard to alternative solutions: when do people truncate their process of investigating an alternative and decide that the status quo is preferable?

Moreover, in real-life Adapt/Exchange decisions, *global problems are often predictable from local problems.* That is, significant costs down the line can be expected as adapting the current solution requires more and more compromises. This accumulation of problems becomes increasingly apparent with time. When will people give up their attempts to adapt? Given the literature on perseveration and adherence to the status quo (Betsch et al., 2002; Jachimowicz et al., 2019; Samuelson & Zeckhauser, 1988; Scherbaum et al., 2013), it can be predicted that this will be rather late. It would be interesting to investigate which factors this depends on, such as the time of problem occurrence or the sunk costs that have already gone into an adaptation.

From an application perspective, future research should contemplate the design of *interventions to support Adapt/Exchange decisions*. Inspiration can be gained from general measures to counteract design fixation (Youmans & Arciszewski, 2014) and specific technologies like case-based reasoning that can enhance creative problem solving (Althuizen & Wierenga, 2014). However, it should be kept in mind that the problem we considered here was not to generate new ideas, but to include them into one's consideration set once they are already known. To this end, assistance systems could reduce information acquisition costs by facilitating the comparison of solutions. Such comparisons should support problem solvers in navigating the abstraction hierarchy. How would changes in a low-level property of a component affect system functions and goal achievement? Or vice versa, given particular functional constraints or goal-related requirements, which low-level changes are compatible with this, and which ones should be avoided? The idea that decision aids should bridge abstraction levels is not new (Bisantz & Vicente, 1994). However, nowadays, model-based assistance systems might support the matching of abstraction levels between human cognitive task representations and formal engineering models (Müller & Urbas, 2023). In this way, a situationally optimal balance between stability and flexibility could be achieved collaboratively by joint cognitive human-machine systems.

# Acknowledgments

We want to thank Thomas Nündel for valuable discussions of mechanical engineering processes, the Adapt/Exchange problems that occur during these processes, and the role of abstraction hierarchies in dealing with them. Parts of this work were funded by a grant of the German Federal Ministry of Education and Research (BMBF, grant number 02K16C070) and a grant of the German Research Foundation (DFG, PA 1232/12-3).